\documentclass[aps,prb,reprint,twocolumn,superscriptaddress,showpacs,floatfix,longbibliography]{revtex4-1}

\usepackage{graphicx}
\usepackage{amsmath}
\usepackage{amssymb}
\usepackage{bbm}
\usepackage{color}
\usepackage{enumitem}
\usepackage{nicefrac}
\usepackage{multirow}
\usepackage{braket}
\usepackage{mathrsfs}

\usepackage[colorlinks,citecolor=blue,urlcolor=blue,bookmarks=false,hypertexnames=true]{hyperref} 

\begin{document}
\title{Improving perturbation theory for open-shell molecules via self-consistency}
\date{\today}
\author{Lan Nguyen Tran}
\email{tnlan@hcmip.vast.vn}
\affiliation{Ho Chi Minh City Institute of Physics, VAST, Ho Chi Minh City 700000, Vietnam}

\begin{abstract}
We present an extension of our one-body M{\o}ller-Plesset second-order perturbation (OBMP2) method for open-shell systems. We derived the OBMP2 Hamiltonian through the canonical transformation followed by the cumulant approximation to reduce many-body operators into one-body ones. The resulting Hamiltonian consists of an uncorrelated Fock (unperturbed Hamiltonian) and a one-body correlation potential (perturbed Hamiltonian) composed of only double excitations. Molecular orbitals and associated energy levels are then relaxed via self-consistency, similar to Hartree-Fock, in the presence of the correlation at the MP2 level. We demonstrate the OBMP2 performance by considering two examples well known for requiring orbital optimization: bond breaking and isotropic hyperfine coupling constants. In contrast to non-iterative MP2, we show that OBMP2 can yield a smooth transition through the unrestriction point and accurately predict isotropic hyperfine coupling constants. 

\end{abstract}

\maketitle

\section{Introduction}
Second-order M{\o}ller-Plesset perturbation theory (MP2) performed on converged Hartree-Fock (HF) orbitals\cite{MP2} may be the simplest correlated wave-function method. Its accuracy depends on the quality of reference wave functions, and it often performs poorly for open-shell systems where unrestricted HF (UHF) orbitals are in many cases spin-contaminated\cite{MP2-JPCA2001, MP2-JCP2011}. 
Moreover, Kurlancheek and Head-Gordon showed that the relaxed MP2 density matrix violates the $N$-representability \cite{MP2-Molphys2009}. This is due to the singularity of the inverse of the orbital Hessian matrix used for evaluating the orbital response of MP2 density matrix.
Recently, orbital-optimized MP2 (OOMP2) and its spin-scaled variants have been developed actively\cite{OOMP2-HeadGordon,OOMP2-Neese,OOMP2-Sherrill,Bozkaya2013-GradOMP2,Bozkaya2014-OMP2,Bozkaya2014-GradOMP2}. In these methods, orbitals are optimized by minimizing the Hylleraas functional (or its generalization) with only double excitations. 
OOMP2 and its variants have been shown to outperform conventional HF-based MP2 calculations for numerous properties. Noticeably, OOMP2 can eliminate the spin contamination present in UHF, providing nearly spin-pure wave functions for open-shell molecules \cite{Bozkaya2014-GradOMP2,OOMP2-HFCC-2010}. Unfortunately, OOMP2 still does not continuously break the spin-symmetry through the unrestriction point. Sharada {\it et. al.} showed that, for OOMP2, the restricted solution is stable with respect to the unrestricted solutions during stretching molecules\cite{OOMP2-Molphys2015}. Therefore, the restricted and unrestricted solutions do not coalesce at the unrestriction point, causing discontinuities in potential energy surfaces \cite{OOMP2-Molphys2015, OOMP2-Molphys2017}. To resolve this issue, Head-Gordon and coworkers have proposed different semi-empirical regularization schemes, including a simple level shift of the energy denominator\cite{OOMP2-JCP2013,OOMP2-Molphys2015,OOMP2-Molphys2017} or orbital energy-dependent regularizers\cite{OOMP2-JCTC2018}.     

In addition to wave-function perturbation theories, double hybrid functionals (DHFs) with a scaled correction of perturbative energy to the solution of Kohn-Sham equations have been actively developed \cite{DHFs-WIRE2014,DHFs-IJC2020}. DHFs are constructed on top of hybrid functionals by replacing part of the DFT correlation functional with a nonlocal correlation contribution based on perturbation theory. They are considered as the fifth rung of the DFT Jacob's ladder and have been shown to outperform conventional functionals in many cases\cite{DHFs-WIRE2014,DHFs-IJC2020}. For example, Kossmann and coworkers found that the functional B2PLYP predicted hyperfine coupling constants more accurately than hybrid functionals and HF-based MP2\cite{DFT-HFCC-2007}. However, similar to HF-based perturbation theory, DHFs also exhibit first derivative discontinuities at the unrestriction point\cite{OODHFs-2013, DHF-JCP2018}. To resolve this artificial problem, Peverati and Head-Gordon proposed orbital-optimized DHFs (OODHFs)\cite{OODHFs-2013}. Much more recently, OOMP2 and OODHF orbitals have been used as references for coupled-cluster singles and doubles with perturbative triples [CCSD(T)] \cite{OOMP2-CC-JCTC2021} and third-order M{\o}ller-Plesset perturbation theory (MP3)\cite{OOMP2-MP3-JPCL2019,OOMP2-OODHF-MP3-JCTC2020}, yielding results several times more accurate than HF-based counterparts. In general, low-level methods generating appropriate reference orbitals for high-level calculations are highly desirable. 

We have developed a self-consistent perturbation theory named one-body MP2 (OBMP2)\cite{OBMP2-JCP2013}. The central idea of OBMP2 is the use of canonical transformation\cite{CT-JCP2006,CT-JCP2007,CT-ACP2007,CT-JCP2009,CT-JCP2010,CT-IRPC2010} followed by the cumulant approximation\cite{cumulant-JCP1997,cumulant-PRA1998,cumulant-CPL1998,cumulant-JCP1999} to derive an effective one-body Hamiltonian. The resulting OBMP2 Hamiltonian is a sum of the Fock operator (unperturbed Hamiltonian) and a one-body correlation potential (perturbed Hamiltonian) including only double excitations. At each iteration, the OBMP2 Hamiltonian, as a {\it correlated} Fock operator, is diagonalized to give eigenvectors and eigenvalues corresponding to molecular orbitals and orbital energies, respectively. Consequently, both numerator (related to molecular orbitals) and denominator (associated with orbital energies) of the double-excitation amplitudes typical for MP2 will be updated simultaneously, resulting in full self-consistency. One can thus expect that OBMP2 can resolve issues caused by the non-iterative nature of standard MP2 calculations.

In the current paper, we extend OBMP2 to open-shell systems. We first present the formulation of OBMP2 in terms of spin orbitals. We then discuss the applications of OBMP2 to two open-shell examples well known for demanding orbital optimization: bond breaking and isotropic hyperfine coupling constants (HFCCs). In contrast to HF-based MP2, we will show that OBMP2 can smooth out the transition from restricted to unrestricted solutions and predict isotropic HFCCs accurately. Finally, we will briefly discuss some implications for OBMP2 extension to a broad class of chemistry problems.

\section{Theory}
We hereafter use the following notation for indices:
$\left\{p, q, r, \ldots \right\}$ refer to general spatial orbitals,
$\left\{i, j, k, \ldots \right\}$ refer to occupied spatial orbitals,
$\left\{a, b, c, \ldots \right\}$ refer to virtual spatial orbitals, and
$\left\{\sigma, \sigma' \right\}$ refer to spin indices.
Einstein's convention is used to present the summations over repeated indices. Our OBMP2 approach was derived through the canonical transformation developed by Yanai and his coworkers\cite{CT-JCP2006,CT-JCP2007,CT-ACP2007,CT-JCP2009,CT-JCP2010,CT-IRPC2010}. In this approach, an effective Hamiltonian that includes dynamic correlation effects is achieved by a similarity transformation of the molecular Hamiltonian $\hat{H}$ using a unitary operator $e^{\hat{A}}$:
\begin{align}
\hat{\bar{H}} = e^{\hat{A}^\dagger} \hat{H} e^{\hat{A}} 
= \hat{H} + \left[\hat{H},\hat{A}\right] + \tfrac{1}{2}\left[\left[\hat{H},\hat{A}\right],\hat{A}\right] + ... \,,
\label{Hamiltonian:ct}
\end{align}
with the anti-Hermitian excitation operator $\hat{A} = - \hat{A}^\dagger$ and the molecular Hamiltonian in spin orbitals as
\begin{align}
  \hat{H} =  h^{p\sigma}_{q\sigma} \hat{a}_{p\sigma}^{q\sigma} + \tfrac{1}{2}g^{p\sigma r\sigma'}_{q\sigma s\sigma'}\hat{a}_{p\sigma r\sigma'}^{q\sigma s\sigma'}\label{eq:h1}
\end{align}
where $h^{p\sigma}_{q\sigma} \hat{a}_{p\sigma}^{q\sigma}$ and $g^{p\sigma r\sigma'}_{q\sigma s\sigma'}\hat{a}_{p\sigma r\sigma'}^{q\sigma s\sigma'}$ are one- and two-electron integrals, respectively. The second equality line in Eq.~\ref{Hamiltonian:ct} is the Baker–Campbell–Hausdorff (BCH) expansion, which is usually cut off at the second-order term. 
In OBMP2, the cluster operator $\hat{A}$ is modeled such that it includes only double excitation
\begin{align}
  \hat{A} = \hat{A}_\text{D} = \tfrac{1}{2}T_{i\sigma j\sigma'}^{a\sigma b\sigma'}(\hat{a}_{i\sigma j\sigma'}^{a\sigma b\sigma'} - \hat{a}_{a\sigma b\sigma'}^{i\sigma j\sigma'}) \,, \label{eq:op1}
\end{align}
with the MP2 amplitude 
\begin{align}
  T_{i\sigma j\sigma'}^{a\sigma b\sigma'} =  \frac{g_{i\sigma j\sigma'}^{a\sigma b\sigma'} } { \epsilon_{i\sigma} + \epsilon_{j\sigma'} - \epsilon_{a\sigma} - \epsilon_{b\sigma'} } \,, \label{eq:amp}
\end{align}
where $\epsilon_{i\sigma}$ is the orbital energy of the spin-orbital $i\sigma$. Substituting the operator $\hat{A}_\text{D}$ into the canonical transformation (Eq~\ref{Hamiltonian:ct}), we can obtain an effective Hamiltonian including dynamical correlation at the level of second-order perturbation. In principle, although one can use a non-unitary exponential operator for the Hamiltonian transformation~(\ref{Hamiltonian:ct}), like in usual coupled cluster theory, using a unitary operator will lead to an effective Hamiltonian that remains Hermitian.        

To derive the working equation of OBMP2 Hamiltonian, we introduce three approximations. First, we truncated the BCH expansion at the second order
\begin{align}
  \hat{H}_\text{OBMP2} = \hat{H}_\text{HF} + \left[\hat{H},\hat{A}_\text{D}\right]_1 + \tfrac{1}{2}\left[\left[\hat{F},\hat{A}_\text{D}\right],\hat{A}_\text{D}\right]_1.
 \label{eq:h2}
\end{align}
Second, we approximate the zeroth-order BCH expansion $\hat{H}$ (Eq.~\ref{eq:h1}) by the HF Hamiltonian $\hat{H}_\text{HF}$
\begin{align}
  \hat{H}_\text{HF} = \hat{F} + C = f^{p\sigma}_{q\sigma} \hat{a}_{p\sigma}^{q\sigma} + \,\, C  \label{eq:h1hf}    
\end{align}
where $\hat{F}$ is the Fock operator, $f^{p\sigma}_{q\sigma}$, and $C$ are the Fock matrix in spin-orbital basis and a constant, respectively:
\begin{align}
f^{p\sigma}_{q\sigma} = & \,\, h^{p\sigma}_{q\sigma} + g^{p\sigma i\sigma'}_{q\sigma i\sigma'} - g^{p\sigma i\sigma'}_{i\sigma' q\sigma} \,, \label{eq:fockhf} \\
C         = & \,\, - g_{i\sigma j\sigma'}^{i\sigma j\sigma'} + g^{i\sigma j\sigma'}_{j\sigma' i\sigma}           \,. \label{eq:consthf}
\end{align}
Third, we employ the cumulant approximation\cite{cumulant-JCP1997,cumulant-PRA1998,cumulant-CPL1998,cumulant-JCP1999} to reduce many-body operators into one-body ones (see Appendix A), such as commutators with the subscription 1, $[\ldots]_1$, involve one-body operators and constants only. 

Substituting Eqs.~\ref{eq:h1}, \ref{eq:op1}, \ref{eq:amp}, \ref{eq:h1hf}, and \ref{eq:fockhf} into Eq.~\ref{eq:h2}, we arrive at the OBMP2 Hamiltonian as follows:
\begin{align}
  \hat{H}_\text{OBMP2} = & \,\, \hat{H}_\text{HF} + \hat{V}_\text{OBMP2} \label{eq:h4} \\
\hat{V}_\text{OBMP2} = & \,\, C' + \hat{V}
\end{align}
with $\hat{V} = v^{p\sigma}_{q\sigma} \hat{a}_{p\sigma}^{q\sigma} $. In the OBMP2 Hamiltonian (Eq.~\ref{eq:h4}), $\hat{H}_{\text{HF}}$ is the unperturbed Hamiltonian and $\hat{V}_{\hat{OBMP2}}$ is the perturbation. $\hat{V}$ and $C'$ consist of contributions from the first and second orders of BCH expansion (Eq.~\ref{eq:h2}): $\hat{V} = \hat{V}_{1^{\text{st}}\text{BCH}} +\hat{V}_{2^{\text{nd}}\text{BCH}}$ and $C' = C'_{1^{\text{st}}\text{BCH}} + C'_{2^{\text{nd}}\text{BCH}}$.
Their working tensor contraction expressions are given as follows:
\begin{widetext}
\begin{align}
\hat{V}_{1^{\text{st}}\text{BCH}} = &  \overline{T}_{i\sigma j\sigma'}^{a\sigma b\sigma'} \left[ f_{a\sigma}^{i\sigma} \,\hat{\Omega}\left( \hat{a}_{j\sigma'}^{b\sigma'} \right) 
  + g_{a\sigma b\sigma'}^{i\sigma p\sigma'} \,\hat{\Omega} \left( \hat{a}_{j\sigma'}^{p\sigma'} \right) - g^{a\sigma q\sigma'}_{i\sigma j\sigma'} \,\hat{\Omega} \left( \hat{a}^{b\sigma'}_{q\sigma'} \right) \right] \nonumber \\
C'_{1^{\text{st}}\text{BCH}} = & - 2 \overline{T}_{i\sigma j\sigma'}^{a\sigma b\sigma'}g^{i\sigma j\sigma'}_{a\sigma b\sigma'}, \nonumber \\ 
\hat{V}_{2^{\text{nd}}\text{BCH}} = 
   & \,f_{a\sigma}^{i\sigma}\overline{T}_{i\sigma j\sigma'}^{a\sigma b\sigma'}\overline{T}_{j\sigma' k\sigma}^{b\sigma' c\sigma} \,\hat{\Omega} \left(\hat{a}_{c\sigma}^{k\sigma} \right)
     +  f_{c\sigma}^{a\sigma}T_{i\sigma j\sigma'}^{a\sigma b\sigma'}\overline{T}_{i\sigma l\sigma'}^{c\sigma b\sigma'} \,\hat{\Omega} \left(\hat{a}^{l\sigma'}_{j\sigma'} \right)               + f_{c\sigma}^{a\sigma}T_{i\sigma j\sigma'}^{a\sigma b\sigma'}\overline{T}_{k\sigma j\sigma'}^{c\sigma b\sigma'} \,\hat{\Omega} \left(\hat{a}^{k\sigma}_{i\sigma} \right) \nonumber \\ 
     &-  f^{k\sigma}_{i\sigma}T_{i\sigma j\sigma'}^{a\sigma b\sigma'}\overline{T}_{k\sigma l\sigma'}^{a\sigma b\sigma'} \,\hat{\Omega} \left(\hat{a}_{l\sigma'}^{j\sigma'} \right)
     -  f^{p\sigma}_{i\sigma}T_{i\sigma j\sigma'}^{a\sigma b\sigma'}\overline{T}_{k\sigma j\sigma'}^{a\sigma b\sigma'} \,\hat{\Omega} \left(\hat{a}^{p\sigma}_{k\sigma} \right)  
     +  f^{k\sigma}_{i\sigma} T_{i\sigma j\sigma'}^{a\sigma b\sigma'}\overline{T}_{k\sigma j\sigma'}^{a\sigma d\sigma'} \,\hat{\Omega}\left(\hat{a}_{b\sigma}^{d\sigma} \right) \nonumber \\
     &+  f_{k\sigma}^{i\sigma}T_{i\sigma j\sigma'}^{a\sigma b\sigma'}\overline{T}_{k\sigma j\sigma'}^{c\sigma b\sigma'} \,\hat{\Omega} \left(\hat{a}_{a\sigma}^{c\sigma} \right) 
     -  f_{c\sigma}^{a\sigma}T_{i\sigma j\sigma'}^{a\sigma b\sigma'}\overline{T}_{i\sigma j\sigma'}^{c\sigma d\sigma'} \,\hat{\Omega} \left(\hat{a}^{b\sigma'}_{d\sigma'} \right) \,
     - f_{p\sigma}^{a\sigma}T_{i\sigma j\sigma'}^{a\sigma b\sigma'}\overline{T}_{i\sigma j\sigma'}^{c\sigma b\sigma'} \,\hat{\Omega} \left(\hat{a}^{p\sigma}_{c\sigma} \right) \nonumber \\
     C'_{2^{\text{nd}}\text{BCH}} = & - 2f_{a\sigma}^{c\sigma}{T}_{i\sigma j\sigma'}^{a\sigma b\sigma'}\overline{T}_{i\sigma j\sigma'}^{c\sigma b\sigma'} +  2f_{i\sigma}^{k\sigma}{T}_{i\sigma j\sigma'}^{a\sigma b\sigma'}\overline{T}_{k\sigma j\sigma}^{a\sigma b\sigma'}, \nonumber
\end{align}
\end{widetext}

where $\overline{T}_{i\sigma j\sigma'}^{a\sigma b\sigma'} = {T}_{i\sigma j\sigma'}^{a\sigma b\sigma'} - {T}_{j\sigma' i\sigma}^{a\sigma b\sigma'}$ and the symmetrization operator $\hat{\Omega} \left( \hat{a}^{p\sigma}_{q\sigma} \right) = \hat{a}^{p\sigma}_{q\sigma}  + \hat{a}^{q\sigma}_{p\sigma}$.
Finally, we rewrite $\hat{H}_\text{OBMP2}$ (Eqs. \ref{eq:h2} and \ref{eq:h4}) in a similar form to Eq. \ref{eq:h1hf} for $\hat{H}_\text{HF}$ as follows:
\begin{align}
  \hat{H}_\text{OBMP2} = & \hat{\bar{F}} + \bar{C} \label{eq:h5}
\end{align}
with $\hat{\bar{F}} =  \bar{f}^{p\sigma}_{q\sigma} \hat{a}_{p\sigma}^{q\sigma}$.
The elements $\bar{f}^{p\sigma}_{q\sigma}$ and $\bar{C}$ are {\it correlated} analogues of the Fock matrix $f^{p\sigma}_{q\sigma}$ [Eq.~\ref{eq:fockhf}] and $C$ [Eq.~\ref{eq:consthf}] of the HF theory and are given as:
\begin{align}
\bar{f}^{p\sigma}_{q\sigma} &= f^{p\sigma}_{q\sigma} + v^{p\sigma}_{q\sigma}, \\
\bar{C}         &= C + C'.                                  \label{eq:fockconstobmp2}
\end{align}
The perturbation matrix $v^{p\sigma}_{q\sigma}$ serves as the correlation potential altering the uncorrelated HF picture. The many-body effective Hamiltonian $\hat{\bar{H}}$ is replaced by the {\it correlated} Fock operator $\hat{\bar{F}}$. The MO coefficients and energies can then be updated by diagonalizing the matrix $\bar{f}^{p}_{q}$, leading to orbital relaxation in the presence of dynamic correlation effects. 

As OBMP2 orbitals are relaxed via full self-consistency similar to HF, OBMP2 satisfies the Brillouin condition: $\bar{f}^{i\sigma}_{a\sigma} = 0$. 
Because the OBMP2 computational cost of each iteration is the same for MP2 scaling [$O(N^5)$], it may serve as a low-cost method generating good reference orbitals for higher-level calculations. For standard MP2, the correlated density matrix is a sum of the uncorrelated HF component and a correction evaluated using the double-excitation amplitude $T_{ij}^{ab}$\cite{OOMP2-HFCC-2010}. In OBMP2, correlation is incorporated into molecular orbitals self-consistently. It is thus worth examining how different molecular properties are evaluated using HF-like and MP2-like OBMP2 density matrices. 
For post-HF methods, one can evaluate the $\left< S^2 \right>$ value by means of either response or projective methods \cite{Bozkaya2014-GradOMP2}. In the response method, $\left< S^2 \right>$ is the sum of the uncorrelated reference value and a correlation correction, and in the projective method, $\left< S^2 \right>$ is the expectation value of the $S^2$ operator. In this paper, to determine how the orbital relaxation under correlation affects the spin contamination, we evaluated $\left< S^2 \right>$ using OBMP2 molecular orbitals (in the same way for UHF). The OBMP2 code is implemented within a local version of PySCF\cite{pyscf-2018}. The next section will demonstrate the applications of OBMP2 to two representative open-shell examples: bond breaking and isotropic HFCCs. 

\section{Results and discussion} \label{sec:iii}
\subsection{Bond breaking}

\begin{figure*}[t!]
  \includegraphics[width=16cm,]{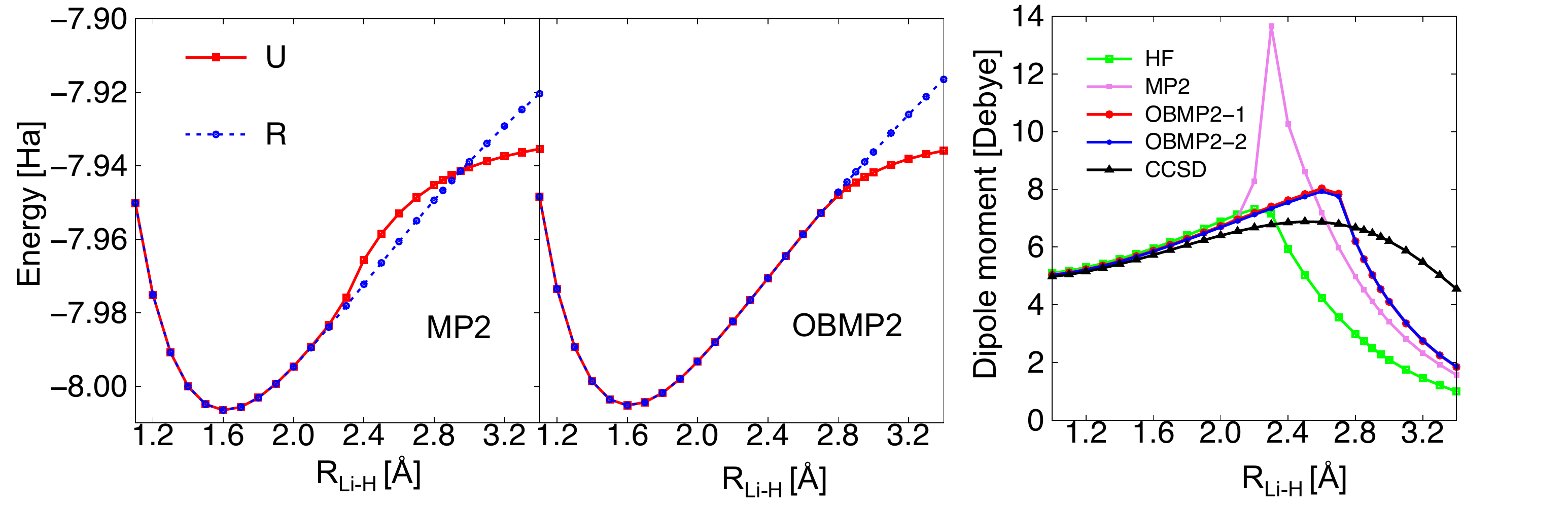}
  \caption{Potential energy curves and dipole moments of LiH in cc-pVDZ. OBMP2 dipole moments are evaluated using HF-like (OBMP2-1) and MP2-like (OBMP2-2) density matrices.}, 
  \label{fig:lih_pes}
\end{figure*}

Previous studies have shown that MP2 often provides incorrect PECs and unphysical discontinuities of first-order properties at the unrestriction point, where the singularity of the inverse of the orbital Hessian matrix used for evaluating the orbital response of MP2 density matrix causes the violation of $N$-representability of relaxed MP2 density matrix\cite{MP2-Molphys2009}. 
Although OOMP2 and its spin-scaled variants are expected to restore the Hellman-Feynman theory and smooth out the transition through the unrestriction point, it was surprising that these methods still often show discontinuities in PECs. To resolve this artifact, one has to employ regularization schemes parameterized semi-empirically \cite{OOMP2-Molphys2015, OOMP2-Molphys2017, OOMP2-JCTC2018}. In OBMP2, orbitals are relaxed in the presence of correlation through full self-consistency; thus, it is worth examining whether OBMP2 can smooth out the transition from restricted to unrestricted solutions. 

Let us start with single-bond breaking: LiH in cc-pVDZ. Figure~\ref{fig:lih_pes} represents the LiH potential energy curves (PEC) and the change of dipole moments with the bond length. We evaluate OBMP2 dipole moments using both HF-like and MP2-like density matrices. For comparison, we also plotted the coupled cluster singles and doubles (CCSD) dipole moment using the unrelaxed density matrix. In contrast to MP2, OBMP2 shows potential energy curves that exhibit a smooth transition from restricted to unrestricted solutions. Both choices of OBMP2 density-matrix evaluation provide nearly identical dipole moments, meaning that the correlation is properly incorporated into molecular orbitals. Unlike MP2, although the OBMP2 still presents a kink at the unrestriction point, it does not have any unphysical jump of the dipole moment. Interestingly, the distance at which the OBMP2 dipole moment starts decreasing is very close to the corresponding CCSD distance.

We now consider the more complicated cases C$_2$H$_4$ and C$_2$H$_2$ that exhibit double- and triple-bond breaking, respectively. Razban {\it et al.} observed that the spin-opposite scaled OOMP2 (SOS-OOMP2) method has stable restricted solutions with respect to unrestricted solutions for these two molecules, resulting in significant discontinuities at unrestriction points\cite{OOMP2-Molphys2017}. They introduced large level shifts to re-balance the restricted and unrestricted solutions such that their orbitals smoothly transform between each other at the unrestriction point. To see the effect of BCH truncation on orbital optimization, we present results from OBMP2 with first-order and second-order BCH truncation. 

\begin{figure}[t!]
  \includegraphics[width=8cm,]{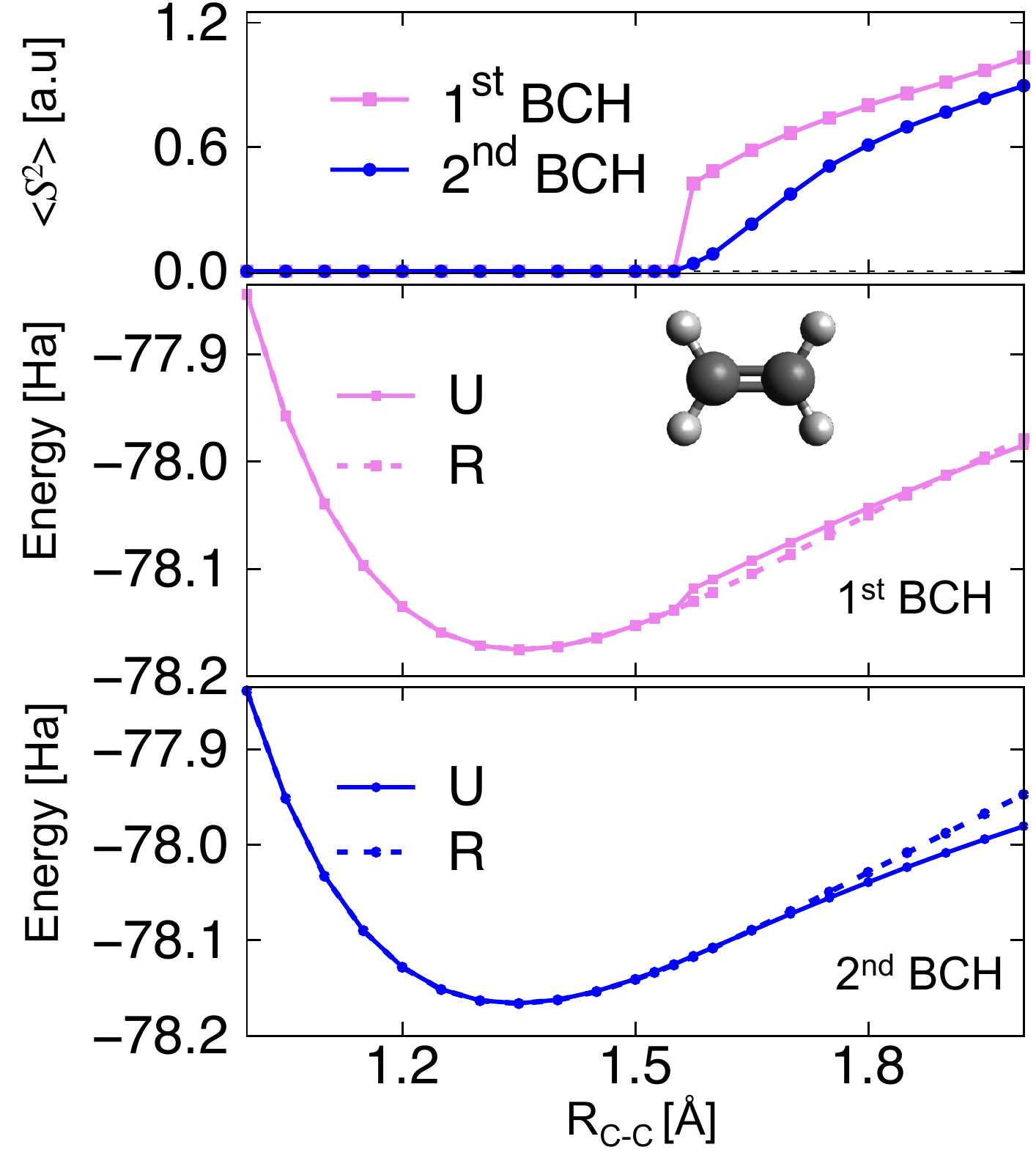}
  \caption{Potential energy curves and $\left<S^2\right>$ of C$_2$H$_4$ in 6-31G from OBMP2.}
  \label{fig:c2h4}
\end{figure}

\begin{figure}[t!]
  \includegraphics[width=8cm,]{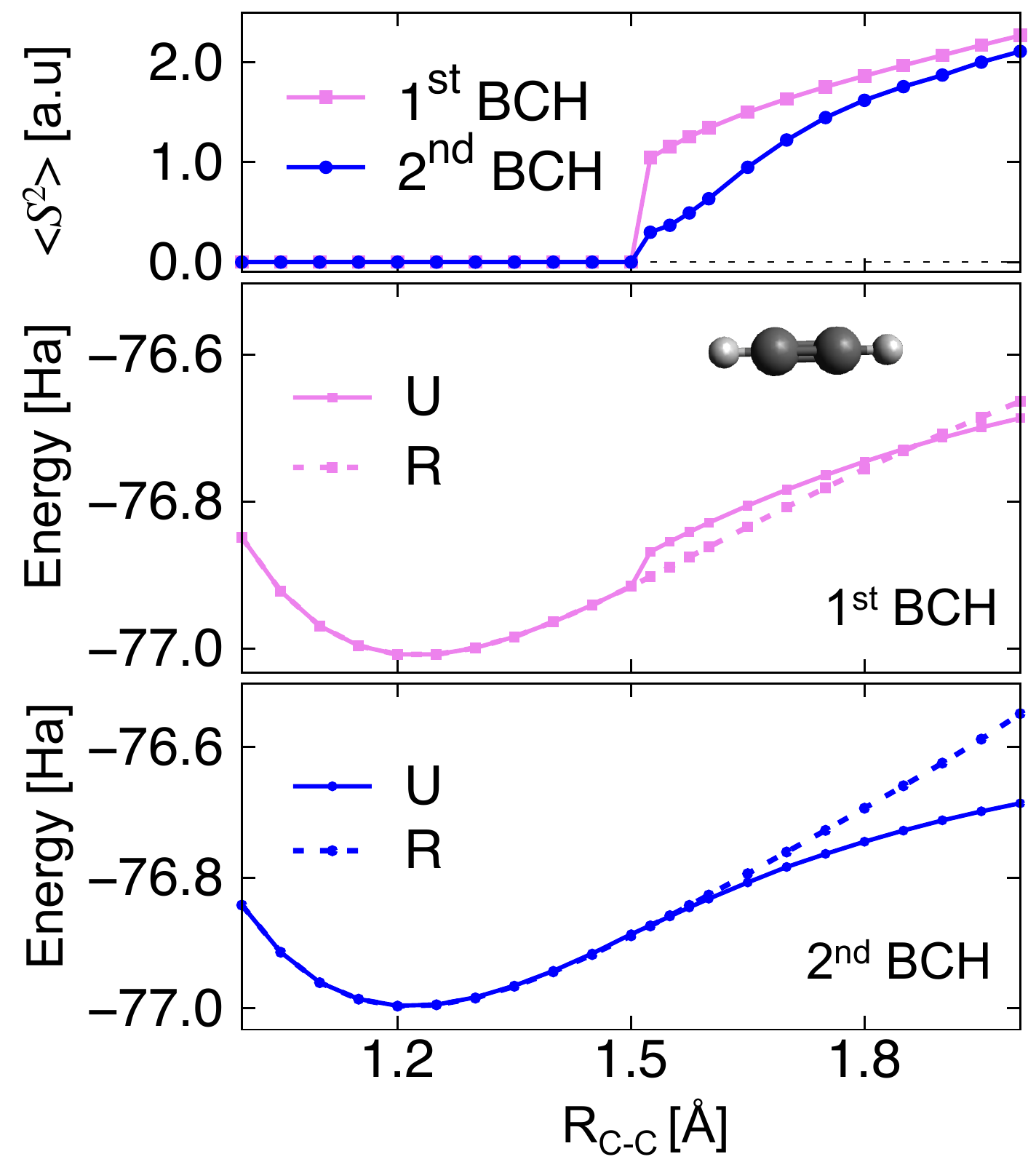}
  \caption{Potential energy curves and $\left<S^2\right>$ of C$_2$H$_2$ in 6-31G from OBMP2.}
  \label{fig:c2h2}
\end{figure}

Figure~\ref{fig:c2h4} represents PECs and the $\left< S^2 \right>$ value of C$_2$H$_4$. We can see that the first-order BCH truncation scheme yields an incorrect PEC with a discontinuity in the unrestricted curves, corresponding to an unphysical jump of $\left< S^2 \right>$ at the unrestriction point. The situation is somewhat similar to the unregularized SOS-OOMP2 results observed by Razban and coworkers\cite{OOMP2-Molphys2017}. In contrast, the second-order BCH exhibits PEC and $\left< S^2 \right>$ that smoothly go through the unrestriction point, meaning that OBMP2 with the second-order BCH truncation scheme can rebalance restricted and unrestricted solutions without the use of any regularization. Figure~\ref{fig:c2h2} represents results for C$_2$H$_2$ that are even more challenging due to the triple-bond breaking. The failure of OBMP2 with the first-order BCH truncation scheme becomes more pronounced. Although the second-order BCH exhibits a small jump of $\left< S^2 \right>$ at the unrestriction point, it can give a smooth PEC.

In general, OBMP2 (with second-order BCH truncation) can yield a smooth transition through the unrestriction point for molecules considered here. This may suggest that the first derivative of OBMP2 energy (nuclear force) is continuous during the molecular dissociation process. However, more extensive testing on a broader test set is necessary. Due to the energy dependence of the denominator of amplitude (Eq.~\ref{eq:amp}), OBMP2 may be drawn into divergence during self-consistency if applying it to systems with small energy gaps. One may need to employ a regularization, such as adding a level shift to the denominator, to restrain the divergence. 

\subsection{Isotropic hyperfine-coupling constants}

We now assess the performance of OBMP2 for the prediction of isotropic HFCCs. The isotropic HFCC describing the interaction between electron and nuclear spins is given by\cite{isotropicHFCC-1988}:  
\begin{align}
  A^K_{iso} = \frac{8\pi}{3} \frac{P_K}{2S} \sum_{\mu \nu} D_{\mu \nu}^{(\alpha-\beta)} \langle \chi_\mu | \delta(r_{K} ) | \chi_\nu \rangle \label{eq:fc}
\end{align}
where $K$ is the index of the given nucleus; $D^{(\alpha-\beta)}$ is the spin-density matrix; $P_K$ is a nucleus-type constant \cite{kaupp2004calculation, OOMP2-HFCC-2010}; $\mu$ and $\nu$ are atomic orbital indices; and $S$ is the total spin. 
It is well known that the accurate prediction of isotropic HFCCs is a highly challenging task for computational quantum chemistry \cite{kaupp2004calculation}. As seen in Eq.~\ref{eq:fc}, the difficulty mainly arises from the direct numerical measure of the spin density at nuclear positions sensitive to the level of the electronic structure method used. 

\begin{table*}[t!]
  \normalsize
  \caption{\label{tab:s2} $\left<S^2\right>$ values from different methods for doublet radicals whose exact value is 0.75.}
  \begin{tabular}{ccccccccccccccccc}
    \hline \hline		
        &UHF$^a$ &MP2$^b$ &OOMP2$^b$	&OBMP2$^a$	&DSD-PBEP86$^c$	&B2PLYP$^c$ \\
    \hline
	BO	&0.7999 &0.7919 &0.7568	&0.7644  &0.7769	&0.7635 \\
    BS	&0.8553	&0.8419 &0.7617 &0.7747  &0.8003	&0.7718 \\
    CO$^+$	&0.9825 &0.9545 &0.7581 &0.7551 &0.8440	&0.7897 \\
    NO	&0.7713 &0.7918 &0.7505	&0.7551  &0.7620	&0.7571 \\
    AlO	&0.8082	&0.7944 &0.7560 &0.7596  &0.7972	&0.7769 \\
    CN	&1.0981	&1.0596 & 0.7524 &0.7658  &0.8701	&0.7868 \\
    CH$_3$	&0.7616 &0.7574 &0.7528	&0.7560   &0.7583	&0.7557 \\
    H$_2$CO &0.7853 &0.7773	&0.7530 &0.7628  &0.7748	&0.7662 \\
    H$_2$O$^+$ &0.7579 &0.7547 &0.7522 &0.7550 &0.7576	&0.7544 \\
    HCO	&0.7656 &0.7598 &0.7513	&0.7557  &0.7599	&0.7573 \\
    MgF	&0.7505 &0.7502 &0.7503	&0.7508  &0.7504	&0.7504 \\
    NH$_2$	&0.7595 &0.7558 &0.7526	&0.7557  &0.7568	&0.7551 \\
    NO$_2$	&0.7709	&0.7629 &0.7503 &0.7547  &0.7636	&0.7595 \\
    OH	&0.7566 &0.7539 &0.7520	&0.7545  &0.7549	&0.7538 \\
    \hline
    MAX &0.3481 &0.0576 &0.0036	&0.0247 &0.1201	&0.0397 \\
    MAD	&0.0660	&0.0586 &0.0035 &0.0086  &0.0305	&0.0142 \\
    RMS &0.1180	&0.0598 &0.0032 &0.0105 &0.0463	&0.0186 \\
    \hline \hline
  \end{tabular}
  \\
   $^a$ Evaluated using molecular orbitals.  \\
   $^b$ Evaluated using the projective method \cite{Bozkaya2014-GradOMP2}. \\
   $^c$ Evaluated from the self-consistent part only (as default in ORCA\cite{orca-2020}).
\end{table*}

With low computational costs, DFT has been widely used for HFCC calculations. Among the functionals tested, hybrid functionals like B3LYP and PBE0 were found to perform best in many cases. However, their success is sometimes attributed to fortuitous error cancellations \cite{kaupp2004calculation}. Some multi-configuration methods, including the density matrix renormalization group (DMRG)\cite{HFCC-DMRG-JCTC2014,HFCC-DMRG-JCTC2015}, complete active space second-order perturbation theory (CASPT2) \cite{HFCC-CASPT2-JCTC2016}, and multi-reference coupled cluster (MRCC) \cite{HFCC-MRCC-JCP2018}, have been assessed for HFCC prediction. Although these methods can provide accurate and reliable HFCCs, they are too expensive for routine applications to large molecules. Recently, Neese and coworkers assessed the performance of double-hybrid functional B2PLYP and (spin-scaled) OOMP2 and found that these methods can reach an accuracy comparable to CCSD(T)\cite{DFT-HFCC-2007,OOMP2-HFCC-2010}. In particular, spin-scaled OOMP2 turned out to be more accurate than standard OOMP2. Let us now assess the performance of OBMP2 (without spin-scaling) for HFCC predictions. Here, we consider a set of 14 small doublet radicals. For comparison, we also evaluated isotropic HFCCs using MP2, two double-hybrid functions B2PLYP and DSD-PBEP86, and CCSD(T) using ORCA\cite{orca-2020}. The EPR-III basis set was employed for all calculations except for the elements Al and Mg, for which the IGLO-III and $s$-decontracted TZVPP basis sets were used, respectively. All geometries were adopted from Ref.~\citenum{gtensor-Neese} except in the case of OH, whose bond length was taken from an experiment. Experimental isotropic HFCCs were adopted from Ref.~\citenum{OOMP2-HFCC-2010}. To see the role of orbital relaxation in OBMP2, we evaluated $D^{(\alpha-\beta)}$ using both HF-like and MP2-like procedures. 

First, to examine whether a method is reliable for the calculations of magnetic properties, we determine the amount of spin contamination in the corresponding calculations. In Table~\ref{tab:s2}, we list $\left< S^2 \right>$ values evaluated from different self-consistent methods: UHF, OBMP2, DSD-PBEP86, and B2PLYP. For comparison, we also present MP2 and OOMP2 $\left< S^2 \right>$ evaluated using the projective method implemented in the Psi4 package\cite{Bozkaya2014-GradOMP2, Psi42017}. Note that, for the double-hybrid functionals, the $\left< S^2 \right>$ value is evaluated only for the self-consistent DFT step by default in ORCA. 
All deviations are given relative to the exact $\left< S^2 \right>$ value of a doublet radical (0.75 a.u.). UHF severely suffers from spin contamination, particularly for CO$^+$ and CN. Although standard MP2 can significantly reduce the spin contamination of the UHF reference, its errors are still large. As previously reported by Bozkaya \cite{Bozkaya2014-GradOMP2}, OOMP2 can mostly eliminate the spin contamination and provide nearly spin-pure wave functions. Both double-hybrid functionals are less spin contaminated than UHF, and B2PLYP performs better than DSD-PBEP86. However, they both still exhibit significant errors for CO$^+$ and CN.  Interestingly, although OBMP2 $\left< S^2 \right>$ is evaluated using molecular orbitals in the same way for UHF, its errors are much smaller than the corresponding HF errors. This means that incorporating correlation into the orbital relaxation part significantly reduces the spin contamination. Noticeably, OBMP2 yields nearly spin-pure wave functions for the two most challenging cases CO$^+$ and CN. Note that the difference between OBMP2 and OOMP2 may mainly originate from different ways of $\left< S^2 \right>$ evaluation. One can evaluate the OBMP2 $\left< S^2 \right>$ value using two-particle density matrices that should give more accurate results than molecular orbitals do.  

\begin{table*}[t!]
  \normalsize
  \caption{\label{tab:hfcs} \normalsize Isotropic hyperfine coupling constant (in MHz) of small doublet radicals. Maximum (MAX) and mean absolute (MAD) deviations relative to experimental values. MP2 and OBMP2 have two isotropic HFCCs evaluated from different density matrices $D$. Experimental istropic HFCCs were adopted from Ref~\onlinecite{OOMP2-HFCC-2010}.}
  \begin{tabular}{cccccccccccccccccccccc}
    \hline \hline		
    \multicolumn{2}{c}{\multirow{2}{*}{Radicals}}  &&\multicolumn{4}{c}{MP2}	&\multicolumn{4}{c}{OBMP2}	&\multirow{2}{*}{DSD-PBEP86}	&&\multirow{2}{*}{B2PLYP}	&&\multirow{2}{*}{CCSD(T)}	&&\multirow{2}{*}{Expt} \\
    \cline{4-6} \cline{8-10}
    & &&unrelaxed  &&relaxed         &&HF-like  &&MP2-like \\
    \hline
BO	&B &&1145.2	&&1002.5	&&1040.4	&&1034.5	&&1022.7	&&1059.7	&&1021.4	&&1033.0 \\
	&O &&23.5	&&--78.2	&&--6.5	&&--6.7	&&--24.8	&&--14.2	&&--31.0	&&--19.0 \\ 
BS	&B &&909.5	&&773.2	&&824.9	&&820.4	&&791.0	&&821.5	&&787.0	&&796.0 \\
	&S &&--23.1	&&39.3	&&1.3	&&2.9	&&9.8	&&5.1	&&--1.9	&&-- \\
CO$^+$	&C &&1901.1	&&1392.6	&&1555.6	&&1540.3	&&1480.2	&&1536.3	&&1513.9	&&1573.0 \\
	&O	&&84.7 &&--99.3	&&35.8	&&30.5	&&0.7	&&23.7	&&--38.9	&&19.0 \\
NO	&N	&&54.6 &&--36.9	&&21.8	&&23.0	&&16.7	&&20.9	&&14.8	&&22.0 \\
	&O	&&--65.2 &&--231.2	&&--30.6	&&--32.3	&&--43.3	&&--34.0	&&--48.6	&&--  \\
AlO	&Al	&&--409.7  &&--75.1	&&843.8	&&860.1	&&1217.2	&&973.6	&&573.2	&&766.0 \\
	&O	&&--53.1 &&128.4	&&17.0	&&9.2	&&64.4	&&21.2	&&9.3	&&2.0 \\
CN	&C	&&1266.1 &&564.5	&&585.5	&&591.8	&&462.0	&&452.5	&&601.8	&&588.0 \\
	&N	&&--34.2 &&15.3	&&--27.9	&&--20.2	&&--22.7	&&--25.6	&&9.6	&&--13.0 \\
CH$_3$	&C	&&161.6 &&58.9	&&83.4	&&83.8	&&77.4	&&84.3	&&73.5	&&75.0 \\
	&H	&&--117.9 &&--71.2	&&--88.8	&&--85.5	&&--74.2	&&--69.8	&&--73.0	&&--70.0 \\
H$_2$CO &H &&236.1	&&282.3	&&316.5	&&316.5	&&297.6	&&325.1	&&298.9	&&372.0 \\
	&C	&&--134.0  &&--94.6	&&--119.3	&&--115.7	&&--97.3	&&--98.8	&&--85.6	&&--109.0 \\
	&O	&&--141.5  &&--42.9	&&--60.4	&&--65.8	&&--55.5	&&--57.3	&&--45.0	&&-- \\
H$_2$O$^+$ &H &&--113.3	&&--74.0	&&--90.3	&&--88.1	&&--75.4	&&--76.8	&&--74.9	&&--73.0 \\
	&O	&&-146.5 &&--68.5	&&--77.4	&&--82.1	&&--79.9	&&--77.4	&&--77.0	&&--83.0 \\
HCO	&H	&&434.0 &&380.6	&&374.1	&&371.8	&&368.3	&&381.1	&&363.4	&&381.0 \\
	&C	&&373.4 &&340.9	&&382.7	&&379.8	&&375.1	&&387.2	&&366.5	&&377.0 \\
	&O	&&-64.8 &&--57.0	&&--40.8	&&--42.4	&&--40.6	&&--39.2	&&--48.7	&&-- \\
MgF	&Mg	&&--273.9   &&--286.6	&&--293.5	&&--292.4	&&--286.8	&&--301.6	&&--282.4	&&--337.0 \\
	&F	&&192.5 &&199.3	&&183.5	&&183.1	&&204.9	&&225.1	&&202.5	&&206.0 \\
NH$_2$	&N &&56.4 &&23.6	&&27.7	&&29.1	&&28.2	&&31.4	&&26.0	&&28.0 \\
	&H	&&--103.6 &&--64.5	&&--79.7	&&--77.1	&&--67.9	&&--66.4	&&--68.7	&&--67.0 \\
NO$_2$	&N &&151.2	&&147.7	&&148.3	&&147.5	&&149.8	&&149.0	&&136.7	&&153.0 \\
	&O	&&--66.5    &&--65.0	&&--62.5	&&--67.6	&&--61.0	&&--62.9	&&--60.0	&&--61.0 \\ 
OH	&O	&&--99.5 &&--40.9	&&--43.2	&&--48.7	&&--48.1	&&--54.2	&&--45.4	&&--51.0 \\ 
	&H	&&-105.1 &&--71.2	&&--84.3	&&--81.8	&&--75.1	&&--73.1	&&--73.5	&&--69.0 \\
	\hline
\multicolumn{10}{l}{Including all cases} \\
\multicolumn{2}{c}{MAX}	&&1175.7 &&841.1	&&77.8	&&94.1	&&451.2	&&207.6	&&192.8 \\
\multicolumn{2}{c}{MAD}	&&127.5 &&67.3	&&16.5	&&16.0	&&37.1	&&24.3	&&23.8 \\
	\hline
\multicolumn{10}{l}{Excluding Al in AlO and C in CN} \\
\multicolumn{2}{c}{MAX}	&&328.1 &&180.4	&&55.5	&&55.5	&&92.8	&&46.9	&&73.1 \\
\multicolumn{2}{c}{MAD}	&&60.8  &&36.8	&&14.6	&&13.3	&&16.1	&&12.0	&&17.2 \\
    \hline \hline
  \end{tabular}
\end{table*}

Table~\ref{tab:hfcs} provides isotropic HFCCs from different theoretical methods and experiments. For HF-based MP2, we evaluated the isotropic HFCCs using both unrelaxed and relaxed density matrices. Here, the MP2 density matrix is relaxed by solving the orbital response (coupled-perturbed HF) equation. Not surprisingly, MP2 with unrelaxed density matrices yields notoriously large errors. Although MP2 with relaxed density matrices reduces errors, they are still huge, meaning that the HF-based MP2 cannot accurately predict isotropic HFCCs. As expected, CCSD(T) can predict isotropic HFCCs well with an acceptable maximum absolute deviation (MAD). However, its maximum absolute deviation (MAX) is still large. For the two DHFs, although B2PLYP yields errors similar to CCSD(T) ones, the newer functional DSD-PBEP8 is less accurate with a huge MAX. The good performance of B2PLYP was previously reported by Neese and coworkers. Looking more closely, the errors of these methods are predominantly due to the Al in AlO, which is well known to be difficult for single-reference methods \cite{AlO-2008, HFCC-DMRG-JCTC2014}. Standard MP2 even predicts the wrong sign for the Al isotropic HFCC. The relaxation of MP2 density matrices via the response equation is insufficient to reach a reasonable accuracy. On the other hand, whereas both DHFs largely overestimate the Al isotropic HFCC, CCSD(T) underestimates it. In addition to the Al HFCC in AlO, the C HFCC in CN is also challenging for the two DHFs that underestimate their isotropic HFCCs. Note that, although standard MP2 with a relaxed density matrix yields a good value for this case, the huge spin contamination of the UHF reference makes this prediction unreliable. If we exclude these two challenging cases, errors of HF-based MP2 and DHFs are reduced significantly. DSD-PBEP86 yields errors comparable to CCSD(T) ones while B2PLYP is even better than CCSD(T).   

\begin{figure*}[t!]
  \includegraphics[width=14cm,]{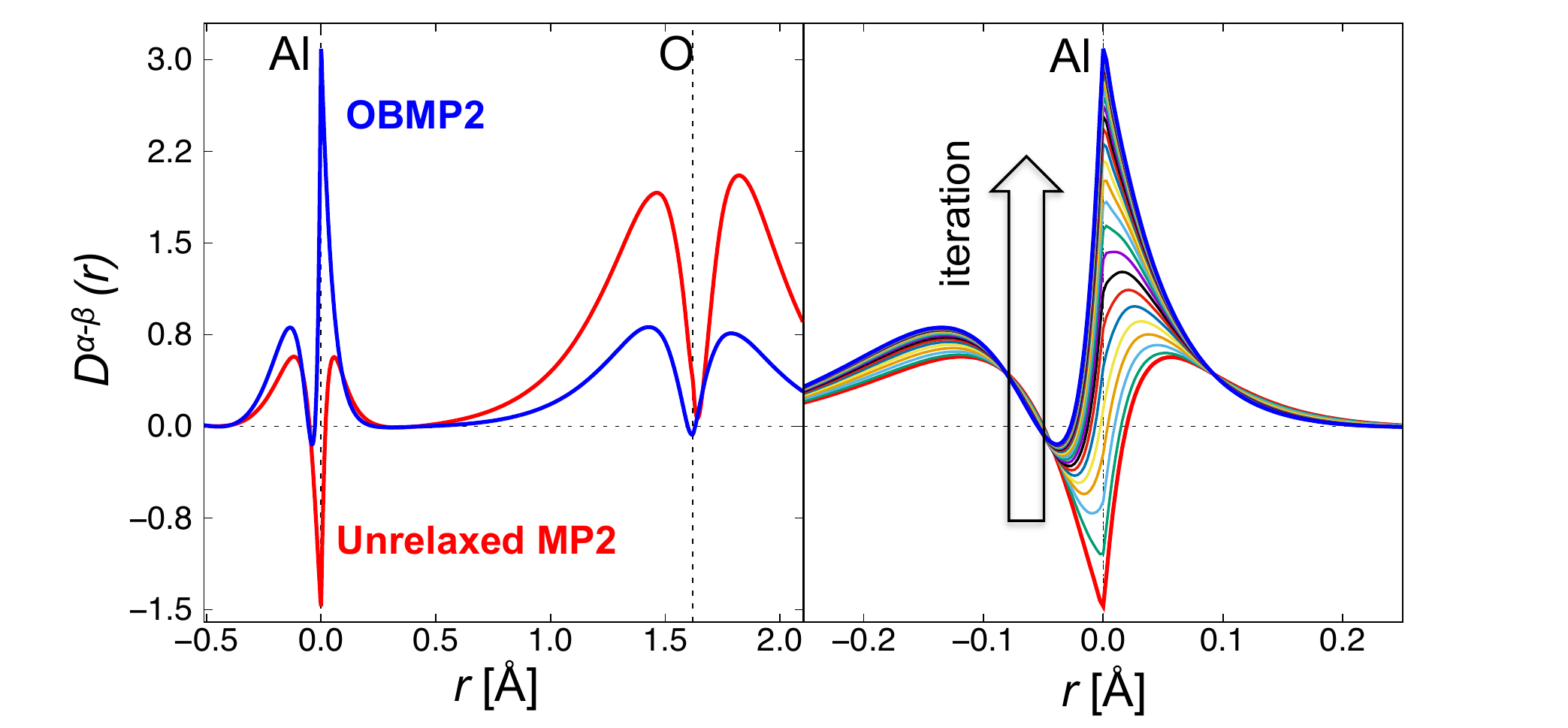}
  \caption{One-dimensional spatial distribution of the AlO spin density of the whole molecule from unrelaxed MP2 and OBMP2 (left) and around the Al center during OBMP2 iterations (right).}
  \label{fig:alo}
\end{figure*}

We now discuss our OBMP2 results. Overall, both approaches to density matrix evaluation (HF-like and MP2-like) give results close to each other, with only a slight deviation. After the first iteration, the OBMP2 isotropic HFCCs using HF-like and MP2-like density matrices are identical to the UHF and unrelaxed MP2 values, respectively. 
OBMP2 can dramatically improve prediction with errors much smaller than MP2 thanks to orbital relaxation via full self-consistency. Although it still overestimates the isotropic HFCC of Al in AlO, its errors are reasonable. Interestingly, OBMP2 is better than CCSD(T) and yields the smallest deviations. Excluding the Al HFCC in AlO and the C HFCC in CN changes the MADs slightly by only a few MHz, meaning that—unlike other methods—OBMP2 can treat these cases well. 

Let us analyze more closely the effect of orbital optimization on isotropic HFCC predictions. Using the most challenging case AlO as an example, we plot in Figure~\ref{fig:alo} the spatial distribution of the MP2-like spin density along the bond axis. In AlO, the singly occupied molecular orbital (SOMO) is characterized by a $\sigma$-bonding between Al($3s$) and O($2p_z$). 
For unrelaxed MP2, whereas the spin density is significantly negative at the Al center, it is largely positive around the O center, explaining why the unrelaxed MP2 gives an overly negative Al isotropic HFCC. In contrast, OBMP2 reduces the spin density around the O center and yields a largely positive one at the Al center, resulting in a reasonable Al isotropic HFCC. From the right panel, we can see that the spin density at the Al center smoothly changes from negative to positive values during the OBMP2 iterations, meaning that orbital relaxation in the presence of correlation is crucial for predicting the isotropic HFCCs accurately. It is important to mention that the spatial distribution of the AlO spin density obtained from OBMP2 has a shape similar to that from DMRG, a much higher-level method (see Figure 2 in Ref.~\citenum{HFCC-DMRG-JCTC2014}). 

In general, from the benchmarking on isotropic HFCC predictions that demands the inclusion of core correlation, we see that OBMP2 is highly promising for accurately predicting magnetic properties. More extensive applications of OBMP2 to other magnetic properties are appealing and essential.  

\section{Implications for OBMP2 extension}\label{sec:iv}

Serving as a fully self-consistent correlated method beyond the uncorrelated HF, OBMP2 has numerous potential applications in chemistry. In addition to thermochemistry and molecular magnetic properties discussed in Section~\ref{sec:iii}, there exist some important possibilities of OBMP2 extension as follows:
\begin{itemize}
    \item {\it Reference orbitals.} As recently shown by Head-Gordon and co-workers, employing OO-DHF and OOMP2 orbitals as references for higher-level calculations like MP3 and CCSD(T) significantly improves the performance for the prediction of many chemistry problems relative to the standard HF-based counterparts. In OBMP2, the correlated potential includes only double excitations, and the orbital relaxation via a full self-consistency guarantees its ground-state wave functions satisfied the Brillouin condition. 
    This suggests that its orbitals can be used in higher-level calculations to improve  their accuracy in particular for open-shell systems. Also, it is interesting to examine whether OBMP2 can serve as a correlated environment (instead of an uncorrelated mean-field) in quantum embedding frameworks, such as density matrix embedding theory \cite{DMET-2012}.
    \item {\it Excited states.} It has been shown that state-specific optimizations are crucial for many excited-state problems like charge transfer and core-level excitation\cite{WGCASSCF-JPCA2020,OODFT-JPCL2021}. Recently, HF-like platforms for excited-state treatment have been developed\cite{ESMF-JCP2018,ESMF-JCP2020}, and dynamic correlation via perturbation theory was incorporated non-iteratively\cite{ESMF-JCP2018,ESPT2-JCTC2020}. However, as we have shown here, orbital relaxation is vital for open-shell systems. Extending OBMP2 to excited states is thus appealing. One can employ the techniques proposed for excited-state DFT like maximum overlap method (MOM)\cite{MOM-JPCA2008} or state-targeted energy projection (STEP)\cite{STEP-JCTC2020} to target an excited state of interest during the OBMP2 iterations.
    \item {\it Perturb-then-diagonalize scheme.} Multi-reference dynamic correlation methods, such as complete active space second-order perturbation theory (CASPT2), are often required for the accurate description of strongly correlated systems. Unfortunately, these methods are costly and limited to small numbers of strongly correlated electrons. Alternatively, one can employ perturb-then-diagonalize methods like non-orthogonal configuration interaction in combination with MP2 (NOCI-MP2)\cite{NOCI-MP2-JCP2016, NOCI-MP2-JCTC2020}, in which single HF determinants constructing Hamiltonian and overlap matrices are corrected by non-iterative MP2. Indeed, it is exciting to use OBMP2 determinants fully relaxed in the presence of dynamic correlation as a basis for these perturb-then-diagonalize schemes.
    \item {\it Double hybrid functionals.} Considered as the fifth rung of Jacob's ladder, DHFs have been actively developed \cite{DHFs-WIRE2014,DHFs-IJC2020}. DHFs mix a portion of correlation energy from perturbation theory, such as MP2 or random phase approximation (RPA), to hybrid exchange-correlation functionals, leading to a better performance than hybrid functionals in many cases. DHFs with orbitals optimized in the presence of all correlations were also developed, resolving some artificial issues present in DHFs with non-iterative correction\cite{OODHFs-2013, OODHF-JPCA2016}. Therefore, it will be interesting to explore new functionals employing OBMP2.
\end{itemize}

\section{Conclusion}

We have developed the unrestricted version of the one-body MP2 (OBMP2) method for open-shell systems. The central idea is to derive an effective one-body Hamiltonian consisting of an uncorrelated Fock and a correlated potential at the MP2 level. Similar to HF counterparts, molecular orbital and orbital energies are then relaxed simultaneously by diagonalizing the correlated Fock matrix. Unlike standard MP2, OBMP2 thus satisfies the Hellmann-Feynamnn theorem, meaning that it can bypass related challenges present in standard MP2. 
The performance of our method has been examined using two representative open-shell examples: bond-breaking and isotropic HFCCs. Our results show that OBMP2 dramatically outperforms standard MP2 for all cases considered here. OBMP2 can exhibit smooth potential energy curves in which restricted and unrestricted solutions coalesce at unrestriction points even for double- and triple-bond breaking in C$_2$H$_4$ and C$_2$H$_2$. A systematic improvement with the order of the BCH expansion was observed. Unlike orbital optimized MP2 and its variants that may need a regularization to smooth out potential energy curves, OBMP2 with the second-order BCH does not require such a semi-empirical procedure for the systems considered here. For isotropic HFCC prediction on a set of main-group doublet radicals, OBMP2 performs better than DFT with double hybrid functionals and CCSD(T), yielding the smallest errors. Plotting the change of spin density during the OBMP2 iterations, we explored the importance of orbital relaxation in the accurate isotropic HFCC prediction. Evaluating OBMP2 $\left< S^2 \right>$ values of those doublet radicals using OBMP2 molecular orbitals in the same way for UHF, we demonstrated that OBMP2 significantly reduces the spin contamination present in UHF and provided nearly spin-pure wave functions. Thus, one can argue that the success of OBMP2 is not fortunate, at least for the test set considered here. 

More extensive assessment is demanded to explore the performance of OBMP2 further. Like standard MP2 and OOMP2, one can use robust techniques, such as density-fitting and Cholesky decomposition \cite{Bozkaya2014-GradOMP2,Bozkaya2014-OMP2,maurer2014cholesky} or local approximations~\cite{werner2003fast,maschio2011local}, to reduce OBMP2 computational costs, making it affordable for large-size applications. As discussed in Section~\ref{sec:iv}, there are many implications for extending OBMP2 to a broader class of chemistry problems. Work on these possibilities is in progress, and we hope to report on them in the future.      

\section*{Appendix}
\subsection*{One-body approximation of many-body operators}
Two-body and three-body operators in the spin-orbital basis, labeled by $\{p, q, r, s, t, u\}$, are approximately written in the one-body form as,
\begin{align}
  \hat{a}^{pr }_{qs } & \Rightarrow  \gamma^{p}_{q} \hat{a}^{r}_{s} + \gamma^{r}_{s} \hat{a}^{p}_{q} - \gamma^{p}_{s} \hat{a}^{r}_{q} - \gamma^{r}_{q} \hat{a}^{p}_{s} - \gamma^{p}_{q} \gamma^{r}_{s} + \gamma^{p}_{s} \gamma^{r}_{q} \label{eq:so-mf2} \nonumber \\
  \hat{a}^{prt}_{qsu} & \Rightarrow  \left( \gamma^{r}_{s} \gamma^{t}_{u} - \gamma^{r}_{u} \gamma^{t}_{s} \right) \hat{a}^{p}_{q} - \left( \gamma^{r}_{q} \gamma^{t}_{u} - \gamma^{r}_{u} \gamma^{t}_{q} \right) \hat{a}^{p}_{s}  \nonumber \\
  &- \left( \gamma^{r}_{s} \gamma^{t}_{q} - \gamma^{r}_{q} \gamma^{t}_{s} \right) \hat{a}^{p}_{u} 
   + \left( \gamma^{p}_{q} \gamma^{t}_{u} - \gamma^{p}_{u} \gamma^{t}_{q} \right) \hat{a}^{r}_{s} \nonumber \\
   &- \left( \gamma^{p}_{s} \gamma^{t}_{u} - \gamma^{p}_{u} \gamma^{t}_{s} \right) \hat{a}^{r}_{q} - \left( \gamma^{p}_{q} \gamma^{t}_{s} - \gamma^{p}_{s} \gamma^{t}_{q} \right) \hat{a}^{r}_{u} \nonumber\nonumber \\
    &+ \left( \gamma^{p}_{q} \gamma^{r}_{s} - \gamma^{p}_{s} \gamma^{r}_{q} \right) \hat{a}^{t}_{u} 
    - \left( \gamma^{p}_{u} \gamma^{r}_{s} - \gamma^{p}_{s} \gamma^{r}_{u} \right) \hat{a}^{t}_{q} \nonumber \\
    &- \left( \gamma^{p}_{q} \gamma^{r}_{u} - \gamma^{p}_{u} \gamma^{r}_{q} \right) \hat{a}^{t}_{s} 
    -2 \left( \gamma^{p}_{q} \gamma^{r}_{s} \gamma^{t}_{u}  - \gamma^{p}_{s} \gamma^{r}_{q} \gamma^{t}_{u} \right) \nonumber \\
    &-2 \left( \gamma^{p}_{u} \gamma^{r}_{s} \gamma^{t}_{q}  + \gamma^{p}_{q} \gamma^{r}_{u} \gamma^{t}_{s} - \gamma^{p}_{s} \gamma^{r}_{u} \gamma^{t}_{q}  - \gamma^{p}_{u} \gamma^{r}_{q} \gamma^{t}_{s} \right)  \nonumber
\end{align}
where
$\hat{a}^{p  }_{q  }$,
$\hat{a}^{pr }_{qs }$, and
$\hat{a}^{prt}_{qsu}$ are the one-, two-, and three-body second-quantized operators, respectively:
$\hat{a}^{p  }_{q  } = \hat{a}^\dagger_{p} \hat{a}_{q}$,
$\hat{a}^{pr }_{qs } = \hat{a}^\dagger_{p} \hat{a}^\dagger_{r} \hat{a}_{s} \hat{a}_{q}$,
$\hat{a}^{prt}_{qsu} = \hat{a}^\dagger_{p} \hat{a}^\dagger_{r} \hat{a}^\dagger_{t} \hat{a}_{u} \hat{a}_{s} \hat{a}_{q}$.
The constant $\gamma^{p  }_{q  }$ is an element of the reduced one-body density matrix, given by:

\begin{align}
 \gamma^{p}_{q} = \langle \Psi_0|\, \hat{a}^{p}_{q} \,|\Psi_0 \rangle \nonumber.
\end{align}

\section*{Acknowledgments}
This work is supported by the Vietnam Academy of Science and Technology (VAST) through the VAST Program for Young Researchers under grant number DLTE00.02/22-23. The authors are also grateful to the HCMC Institute of Physics, VAST, for encouragement and support. We performed all calculations on the High-Performance Computing system located at the Center for Informatics and Computing (CIC), VAST.  
\bibliography{main}
\end{document}